# Stacking-dependent spin interactions in Pd/Fe bilayers on Re(0001)

W. Li[1,*], S. Paul[2,*], K. von Bergmann[1], S. Heinze[2], and R. Wiesendanger[1,#]

[1]Department of Physics, University of Hamburg, Jungiusstrasse 9-11, 20355 Hamburg, Germany
[2]Institute of Theoretical Physics and Astrophysics, University of Kiel, Leibnizstrasse 15, 24098 Kiel, Germany

Abstract:

Using spin-polarized scanning tunneling microscopy and density functional theory, we have studied the magnetic properties of Pd/Fe atomic bilayers on Re(0001). Two kinds of magnetic ground states are discovered due to different types of stacking of the Pd adlayer on Fe/Re(0001). For fcc-stacking of Pd on Fe/Re(0001), it is a spin spiral propagating along the close-packed ($\overline{\Gamma K}$) direction with a period of about 0.9 nm, driven by frustrated exchange and Dzyaloshinskii-Moriya interactions. For the hcp stacking, the higher-order exchange interactions stabilize an *uudd* (up-up-down-down) state propagating perpendicular to the close-packed direction (along $\overline{\Gamma M}$) with a period of about 1.0 nm.

Recently, non-collinear magnetic states, such as spin spirals, skyrmions and multi-Q states, have caused a lot of attention as model-type systems to investigate the exciting physics of complex spin interactions, including higher-order and multi-spin interactions, and as promising platforms for the development of future magnetic memory and logic devices based on topological magnetic quasiparticles [1,2]. In parallel, topological superconductors (TSCs) and associated Majorana modes, have become very hot topics because of their potential application in quantum computation [3]. According to theoretical predictions, hybrid systems of magnetic nanostructures and s-wave superconductors could be a very promising platform to realize Majorana modes [4,5]. Experimentally, signatures of Majorana modes have been observed at the ends of ferromagnetic [6] and spin spiral atomic chains [7] as well as at the edge of magnetic nanoislands [8] interacting with s-wave superconductors using spin-polarized scanning tunneling spectroscopy (SP-STS) [9]. To explore more possible systems for the realization of Majorana modes, different kinds of spin textures on superconductors have been proposed by theory [10,11]. Therefore, the investigation of the magnetic order of adlayers on superconductors and related physics has become a focus of current research activities. Some theoretical studies even proposed to combine magnetic skyrmions with superconductors to manipulate Majorana states [12-14].

Pd/Fe atomic bilayers on Ir(111) is a well-studied system exhibiting a spin spiral ground state in zero field and magnetic skyrmions in an applied external magnetic field as revealed by spin-polarized scanning tunneling microscopy (SP-STM) [15,16]. More importantly, it is the first reported system used for writing and deleting of single skyrmions by a SP-STM tip [15]. However, the superconducting transition temperature $T_c$ of an Ir substrate (~ 0.11 K) is too low to observe clearly zero-energy Majorana states within the energy gap of the superconductor by scanning tunneling spectroscopy (STS). An ideal replacement for the Ir(111) substrate is Re(0001), which has a higher $T_c$ of about 1.7 K and a surface lattice constant (~ 0.276 nm) close to the one of Ir(111) (~ 0.272 nm).

Motivated by the future use of Pd/Fe bilayers on Re(0001) as a possible model-type platform to realize topological superconductivity, we report here on the epitaxial growth as well as on the investigation of the resulting spin textures of Pd/Fe/Re(0001) by a combined SP-STM and density functional theory (DFT) study. The experimental findings are compared with the results of the DFT calculations and reveal a surprising dependence of the observed spin texture on the type of stacking of the Pd top layer. In particular, the four-site four spin interaction changes sign due to the modified hybridization at the Pd/Fe/Re interface in fcc vs. hcp stacking which triggers the change of the magnetic ground state.

Fig. 1(a) shows an STM topography of pseudomorphic Fe islands on clean terraces of the Re(0001) substrate as well as Fe nanostripes close to the Re(0001) step edges after depositing a sub-monolayer of Fe on Re(0001) held at room temperature. According to previous reports [17-19], the stacking of Fe on Re(0001) should be hcp-type. The inset of Fig. 1(a) shows a zoom-in revealing the atomic lattices of an Fe nanostripe and an upper-terrace of bare Re(0001) simultaneously. From such images, the hcp stacking of Fe on Re(0001) can be confirmed by our present experiment. Pd was then deposited onto the Fe/Re(0001) sample held at room temperature. Since the Re(0001) substrate is not completely covered by the monolayer Fe film, several different kinds of layers are formed including monolayer Pd on Re(0001), double-layer Pd on Re(0001), monolayer Pd on Fe/Re(0001), and double-layer Pd on Fe/Re(0001). By comparing the STM topography image of Fig. 1(b) and the simultaneously recorded differential tunneling conductance (*dI/dU*) map (Fig. 1(c)) at some favorable sample bias (here: +200 mV), the different layers exposed can easily be distinguished. Figure 1(d) shows *dI/dU* spectra of the different layers, indicating at which particular sample bias a high *dI/dU* contrast between the different layers can be expected. For instance, by using a sample bias of +200 mV, the monolayer Pd on Re(0001) has the highest *dI/dU* intensity, followed by monolayer Pd on Fe/Re(0001) and double-layer Pd on Re(0001), while the double-layer Pd on Fe/Re(0001) has the lowest *dI/dU* intensity, in agreement with the various contrast levels observed in Fig. 1(c). The exposed surfaces of Fe/Re(0001) and Re(0001) can very easily be distinguished because of their rough STM topography which results from hydrogen adsorption during the Pd evaporation.

The hexagonally close-packed Pd adlayer on hcp-Fe/Re(0001) (see inset of Fig. 1(b)) can, in principle, have two different types of stacking, i.e., fcc or hcp. Usually, adlayers with different stacking exhibit a similar STM topography, but different contrast in *dI/dU* maps. As evident from Fig. 2(a), the marked Pd islands all show a similar topography, but they exhibit two different contrast levels in the simultaneously recorded *dI/dU* map at a sample bias of -100 mV (Fig. 2(b)): one type of island appears darker (corresponding to a low *dI/dU* intensity), while the other island type appears brighter (corresponding to a higher *dI/dU* intensity). This difference in the *dI/dU* contrast is attributed to the two different types of stacking which can occur, i.e. fcc- and hcp-Pd adlayers on hcp-Fe/Re(0001). As shown in the supplementary information (Fig. S1), the darker and brighter islands refer to fcc- and hcp-Pd on hcp-Fe/Re(0001), respectively. To study the magnetic properties of both fcc- and hcp- islands of Pd/Fe/Re(0001), spin-resolved *dI/dU* measurements were performed for the same surface region using a spin-sensitive Cr bulk tip. In the magnetic *dI/dU* map of Fig. 2(c), which was measured without an external magnetic field applied, periodic stripes are observed on the Pd/Fe/Re(0001) islands, which are not present in *dI/dU* images obtained with non-magnetic STM probe tips. Interestingly we found that the stripes on fcc- and hcp-Pd/Fe/Re(0001) islands have different crystallographic directions and periods. On fcc-Pd islands, the stripes are perdendicular to the close-packed direction ($[1\bar{1}0]$, as marked in Fig. 2(c-e)) with a period of 0.9 nm (see Fig. 2(f)), while on hcp-Pd islands, the stripes are parallel to the $[1\bar{1}0]$ direction with a period of 1.0 nm (see Fig. 2(g)).

To ultimately prove that the stripes in the spin-resolved *dI/dU* maps originate from a spin texture rather than from a collective electronic state (e.g. a charge density wave), measurements with a different magnetic moment orientation of the SP-STM probe tip were performed. By making use of different micro-tips with varying directions of the apex magnetic moment (see supplementary Fig. S2), we finally establish the existence of two spin textures with different periods and propagation directions originating from two distinguishable magnetic ground states.

To understand the reason for the two different magnetic ground states observed experimentally, depending on the stacking sequence of Pd on Fe/Re(0001), we have investigated the structural, electronic and magnetic properties of Pd/Fe bilayers on Re(0001) using DFT. In order to scan a large part of the magnetic phase space, we have calculated the energy dispersion of flat homogeneous spin spirals without spin-orbit coupling (SOC) along the two high-symmetry directions of the two-dimensional Brillouin zone (2DBZ) for hcp-Pd/Fe/Re(0001) and fcc-Pd/Fe/Re(0001) (light color plots in Fig. 3 is without SOC). For both stackings of Pd, we observe that spin spiral states along $\overline{\Gamma K}$ and $\overline{\Gamma M}$ directions have lower energy than the ferromagnetic (FM) state ($\bar{\Gamma}$ point). However, the spin spirals along $\overline{\Gamma K}$ have the lowest energy for both stackings. For fcc-Pd/Fe/Re(0001), the spin spiral energy minimum is ∼3.5 meV/Fe atom, and for hcp-Pd/Fe/Re(0001), it is ∼24 meV/Fe atom lower than the FM state. The magnetic moments of Fe in fcc-Pd/Fe/Re(0001) and hcp-Pd/Fe/Re(0001) are about 2.3 $\mu_B$ and 2.6 $\mu_B$, respectively, and vary little with **q** (see Supplementary Fig. S3). The induced magnetic moments in Pd and Re amount to up to 0.4 $\mu_B$ and −0.2 $\mu_B$, respectively. We calculate the exchange constants by mapping the spin spiral energies without SOC onto the Heisenberg model (see supplementary table S3). We find that the nearest-neighbor exchange interaction is FM, whereas the second and third nearest-neighbor exchange interactions are antiferromagnetic (AFM), implying a frustration of exchange interactions, which stabilize the spin spiral energy minima.

Including SOC, the energy difference between the FM state and the spin spiral state minimum along $\overline{\Gamma K}$ increases and we find that it is ∼10 meV/Fe atom ($\lambda$ = 1.1 nm) for fcc-Pd/Fe/Re(0001) and ∼29 meV/Fe atom ($\lambda$ = 0.92 nm) for hcp-Pd/Fe/Re(0001) lower than the FM state. We attribute this to the DMI, which favors right-rotating spin spirals for both stackings (see Supplementary Fig. S4). Note that the experimental period of the magnetic ground state for fcc-Pd/Fe/Re(0001) is about 0.9 nm, which matches quite well with the value obtained by DFT. However, based on spin spiral calculations – even including SOC – it is not possible to understand why the experimental data shows a spin structure propagating along the $\overline{\Gamma K}$ direction for fcc-Pd stacking, while it propagates along $\overline{\Gamma M}$ for hcp-Pd stacking. Therefore, we study the effect of higher-order exchange interactions (HOI), which can lead to complex magnetic ground states as has been shown for ultrathin films, such as Fe/Ir(111) [20], Rh/Fe/Ir(111) [21] and Fe/Rh(111) [22]. To investigate the effect of HOI, we calculate three multi-*Q* states: (i) two collinear spin structures along $\overline{\Gamma K}$ and $\overline{\Gamma M}$, the so-called *uudd* state or double-row wise AFM state [23]; and (ii) a three-dimensionally modulated non-collinear spin state at the $\bar{M}$ point, the so-called 3*Q* state [24], for both stackings of Pd. The multi-*Q* states are superpositions of spin spirals corresponding to symmetry-equivalent **q** vectors in the 2DBZ. Therefore, by construction, these states are energetically degenerate with the corresponding spin spiral states (single-$Q$ states) within the Heisenberg model. However, this degeneracy is lifted by HOI which can be quantified from the energy difference between the multi-Q and single-Q states (see supplementary table S3).

The energies of all multi-$Q$ states are shown in Fig. 3. For fcc-Pd/Fe/Re(0001), we find that all three multi-$Q$ states are energetically higher than the corresponding spin spiral states, i.e., the aforementioned spin spiral state remains the magnetic ground state. On the other hand, all three multi-$Q$ states for hcp-Pd/Fe/Re(0001) are energetically lower than the corresponding spin spiral states. Most importantly, the *uudd* state along $\overline{\Gamma M}$ is ~3 meV/Fe atom lower in energy than the lowest spin spiral state. Therefore, the ground state of hcp-Pd/Fe/Re(0001) is the *uudd* state with a period of 0.96 nm, which matches quite well with the experimental observation of a 1.0 nm period.

The energy differences of the multi-$Q$ states with respect to the spin spiral states demonstrate the influence of the stacking order of the Pd overlayer on the higher-order exchange interactions. We find that the HOI are dominated by the four-site four spin interaction (see Supplementary Table S3). The transition from the spin spiral to the *uudd* ground state is directly related to the change of sign of the four-site four spin interaction from fcc-Pd to hcp-Pd stacking. Previously, it has been demonstrated that the four-site four spin interaction is responsible for the nanoskyrmion lattice of Fe/Ir(111) [20] and that it plays an important role for the stability of isolated skyrmions [26]. In contrast, it is the three-site four spin interaction which stabilizes the *uudd* state in the Fe monolayer on Rh(111) [22].

To relate the stacking dependent magnetic ground state to the electronic structure, we have calculated the local density of states (LDOS) of the topmost three layers in the FM state for fcc-Pd/Fe/Re(0001) and hcp-Pd/Fe/Re(0001) [Fig. 4]. We observe significant changes of the LDOS in the majority and minority spin channels in all three layers upon changing the stacking of the Pd layer, which indicates a large modification of hybridization at the interface. Prominent changes in the Fe majority spin channel [Fig. 4(b)] are three peaks in an energy range of 1 eV below the Fermi level, $E_F$, for hcp-Pd stacking which move to lower energy for fcc stacking. In the minority channel of Fe, the two peaks just below $E_F$ of hcp-Pd stacking shift towards $E_F$ for fcc-Pd/Fe/Re(0001).

As a result, the spin-polarization at $E_F$ is enhanced for fcc-Pd and the Fermi surface is modified, which affects the exchange interactions (Supplementary Table S3). The four-site four-spin exchange interaction arises due to cyclic hopping of electrons on sites along a diamond-shaped path and its strength also depends on the orbital wave functions. Therefore, the stacking order and modified electronic structure also affects these higher-order interactions.

Usually, magnetic states can be modified, and even be tuned from topologically trivial to non-trivial (e.g., spin spirals to skyrmions [15,25]) by an external magnetic field. Therefore, we have performed magnetic-field dependent studies of both the fcc- and hcp-stacked Pd on Fe/Re(0001) islands. The measured *dI/dU* maps at 5 T (Fig. 2(d)) and at 9 T (Fig. 2(e)) appear similar to the measurement performed at zero magnetic field (Fig. 2(c)). Obviously, both magnetic ground states, i.e., spin spiral and *uudd* state, do not respond to an external magnetic field of up to 9 T. These experimental observations are consistent with the DFT results of Fig. 3. The spin spiral ground state of fcc-

Pd/Fe/Re(0001) is nearly 10 meV lower than the FM state, which requires a magnetic field of ~17 T to stabilize skyrmions and for hcp-stacked Pd, the critical field is even larger.

In conclusion, by performing SP-STM/STS measurements combined with DFT calculations, we studied the electronic and magnetic properties of Pd/Fe bilayers on Re(0001) and discovered two different stacking-dependent magnetic ground states. For fcc-Pd/Fe/Re(0001), the concerted interplay of frustrated exchange and DMI stabilizes a right-rotating cycloidal spin spiral along the $\overline{\Gamma K}$ direction, whereas for hcp-Pd/Fe/Re(0001) the four-site four-spin interactions favor a *uudd* state along the $\overline{\Gamma M}$ direction. Our results establish the Pd/Fe/Re(0001) system as a promising platform for studying the interaction of complex spin textures with a superconducting substrate, potentially leading to novel types of exotic states such as topological superconductivity in the hybridized system

We would like to thank André Kubetzka for useful discussions and technical assistance for the experiments. We gratefully acknowledge financial support from the European Union via the ERC Advanced Grant ADMIRE (project No. 786020). K.v.B. acknowledges financial support from the Deutsche Forschungsgemeinschaft (DFG, German Research Foundation) -418425860; -402843438. S. H. thanks the Deutsche Forschungsgemeinschaft (DFG, German Research Foundation) for funding via project no. -418425860.

References:

[*] These authors contributed equally to this work.
[#] wiesendanger@physnet.uni-hamburg.de
[1]  F. Hellman *et al.*, Rev. Mod. Phys. **89**, 025006 (2017).
[2]  R. Wiesendanger, Nat. Revs. Mater. **1**, 16044 (2016).
[3]  X.-L. Qi and S.-C. Zhang, Rev. Mod. Phys. **83**, 1057 (2011).
[4]  B. Braunecker and P. Simon, Phys. Rev. Lett. **111**, 147202 (2013).
[5]  J. Klinovaja, P. Stano, A. Yazdani, and D. Loss, Phys. Rev. Lett. **111**, 186805 (2013).
[6]  S. Nadj-Perge, I. K. Drozdov, J. Li, H. Chen, S. Jeon, J. Seo, A. H. MacDonald, B. A. Bernevig, and A. Yazdani, Science **346**, 602 (2014).
[7]  H. Kim, A. Palacio-Morales, T. Posske, L. Rozsa, K. Palotas, L. Szunyogh, M. Thorwart, and R. Wiesendanger, Sci. Adv. **4**, eaar5251 (2018).
[8]  A. Palacio-Morales, E. Mascot, S. Cocklin, H. Kim, S. Rachel, D. K. Morr, and R. Wiesendanger, Sci. Adv. **5**, eaav6600 (2019).
[9]  R. Wiesendanger, Rev. Mod. Phys. **81**, 1495 (2009).
[10] S. Nakosai, Y. Tanaka, and N. Nagaosa, Phys. Rev. B **88**, 180503(R) (2013).
[11] J. Rontynen and T. Ojanen, Phys. Rev. Lett. **114**, 236803 (2015).
[12] G. Yang, P. Stano, J. Klinovaja, and D. Loss, Phys. Rev. B **93**, 224505 (2016).
[13] U. Güngördü, S. Sandhoefner, and A. A. Kovalev, Phys. Rev. B **97**, 115136 (2018).
[14] S. Rex, I. V. Gornyi, and A. D. Mirlin, Phys. Rev. B **100**, 064504 (2019).
[15] N. Romming, C. Hanneken, M. Menzel, J. E. Bickel, B. Wolter, K. von Bergmann, A. Kubetzka, and R. Wiesendanger, Science **341**, 636 (2013).
[16] N. Romming, A. Kubetzka, C. Hanneken, K. von Bergmann, and R. Wiesendanger, Phys. Rev. Lett. **114**, 177203 (2015).

[17] S. Ouazi, A. Kubetzka, K. von Bergmann, and R. Wiesendanger, Phys. Rev. Lett. **112**, 076102 (2014).
[18] S. Ouazi, T. Pohlmann, A. Kubetzka, K. von Bergmann, and R. Wiesendanger, Surf. Sci. **630**, 280 (2014).
[19] A. Palacio-Morales, A. Kubetzka, K. von Bergmann, and R. Wiesendanger, Nano Lett. **16**, 6252 (2016).
[20] S. Heinze, K. von Bergmann, M. Menzel, J. Brede, A. Kubetzka, R. Wiesendanger, G. Bihlmayer, and S. Blügel, Nat. Phys. **7**, 713 (2011).
[21] N. Romming, H. Pralow, A. Kubetzka, M. Hoffmann, S. von Malottki, S. Meyer, B. Dupé, R. Wiesendanger, K. von Bergmann, and S. Heinze, Phys. Rev. Lett. **120**, 207201 (2018).
[22] A. Krönlein *et al.*, Phys. Rev. Lett. **120**, 207202 (2018).
[23] B. Hardrat, A. Al-Zubi, P. Ferriani, S. Blügel, G. Bihlmayer, and S. Heinze, Phys. Rev. B **79**, 094411 (2009).
[24] P. Kurz, G. Bihlmayer, K. Hirai, and S. Blügel, Phys. Rev. Lett. **86**, 1106 (2001).
[25] M. Herve, B. Dupe, R. Lopes, M. Bottcher, M. D. Martins, T. Balashov, L. Gerhard, J. Sinova, and W. Wulfhekel, Nat. Commun. **9**, 1015 (2018).
[26] S. Paul, S. Haldar, S. von Malottki, and S. Heinze, arXiv:1912.03474 [cond-mat.mes-hall] (2019).

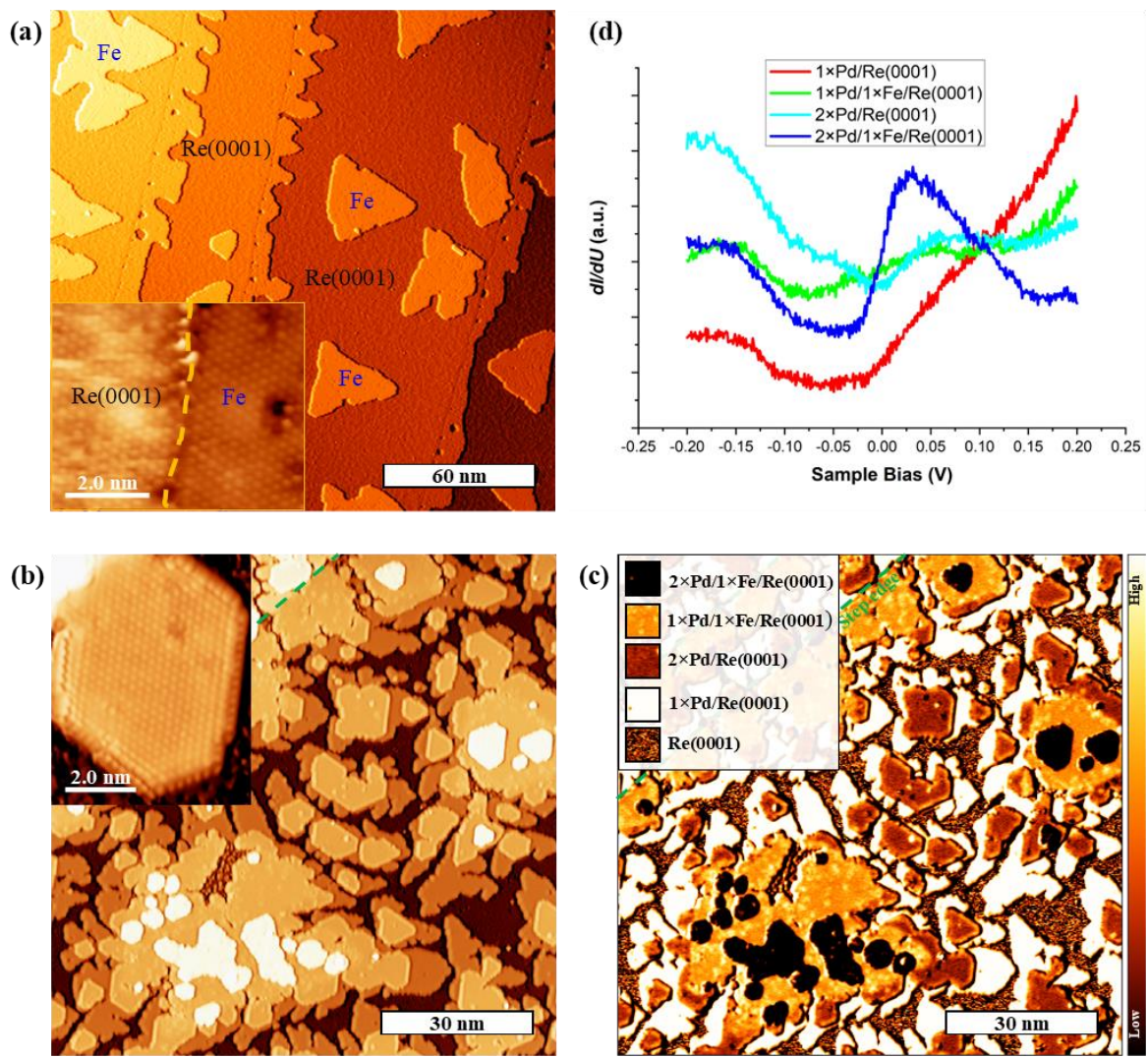

**Figure 1**. (a) STM topography of pseudomorphic Fe monolayer (ML) on Re(0001). The coverage is about 0.3 ML ($U$ = 100 mV, $I$ = 1 nA). The inset is a zoom-in on both the Re(0001) substrate and the Fe monolayer close to a Re(0001) step edge (marked by a yellow dashed line), indicating an hcp-stacking of Fe on Re(0001) ($U$ = 5 mV, $I$ = 20 nA). (b) STM topography of Pd on Fe/Re(0001). The Fe coverage is about 0.3 ML, and the Pd coverage is about 1 ML ($U$ = 200 mV, $I$ = 2 nA). The inset is a zoom-in on a Pd island on Fe/Re(0001) ($U$ = 20 mV, $I$ = 2 nA). (c) Simultaneously recorded $dI/dU$ map of (b) allowing the distinction between different layers exposed. (d) Point $dI/dU$ spectra of the different layers.

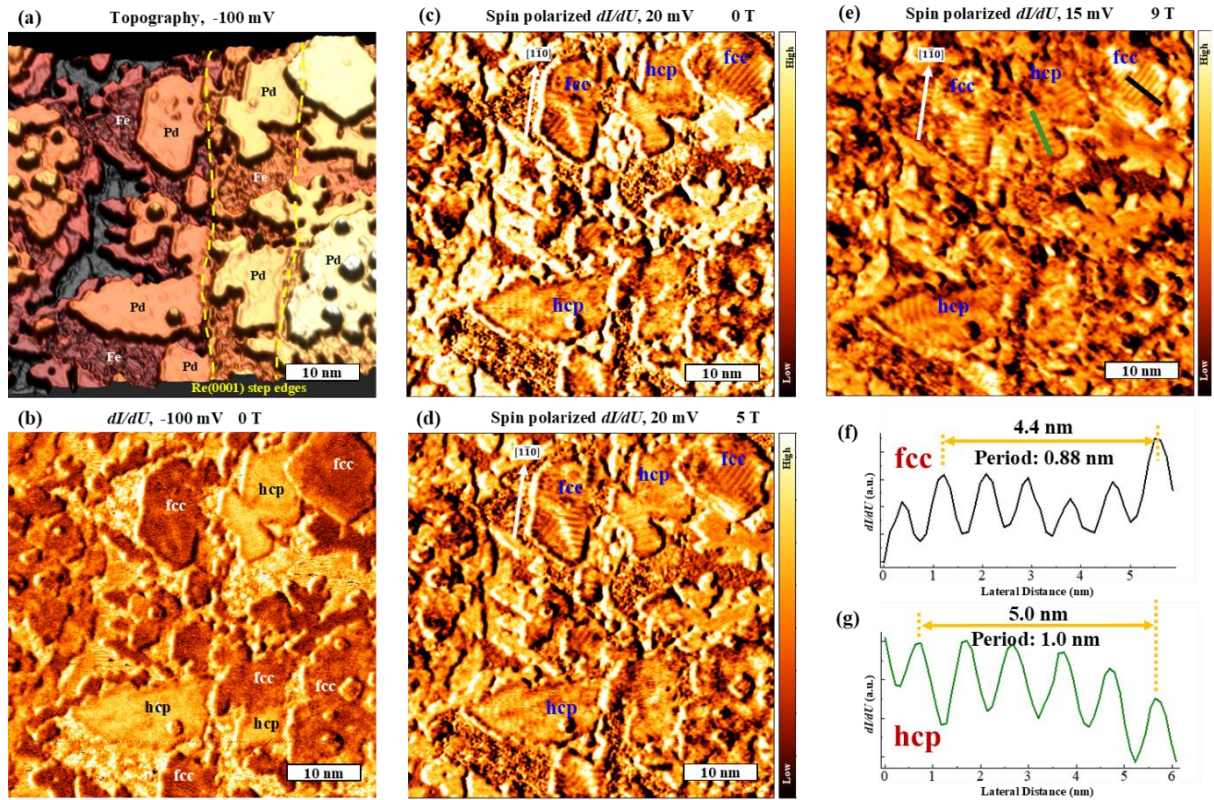

**Figure 2**. (a) STM topography and (b) simultaneously recorded *dI/dU* map (*U* = -100 mV, *I* = 2 nA, 0 T, Cr bulk tip). The monolayer Pd islands on Fe/Re(0001) are marked in (a). The Pd/Fe/Re(0001) islands show two different contrast levels in the *dI/dU* map at *U* = -100 mV. This is attributed to two different stacking types, i.e. fcc and hcp of Pd on Fe/Re(0001). (c-e) Spin-polarized *dI/dU* maps at 0T, 5T and 9T out-of-plane magnetic field. (*U* = 15-20 mV, *I* = 2 nA, Cr bulk tip). Periodic stripes can be observed on the Pd/Fe/Re(0001) islands in the spin-polarized *dI/dU* maps with a different contrast for fcc- and hcp-stacked islands: The stripe period for fcc islands is about 0.9 nm, and the stripe directions are perpendicular to the close-packed directions of the islands. For hcp islands, the observed period is about 1.0 nm, and the stripe directions are along the close-packed directions of the islands. (f-g) line profiles of the periodic stripes in fcc- and hcp- stacked Pd/Fe/Re(0001) islands (marked in (e)), respectively.

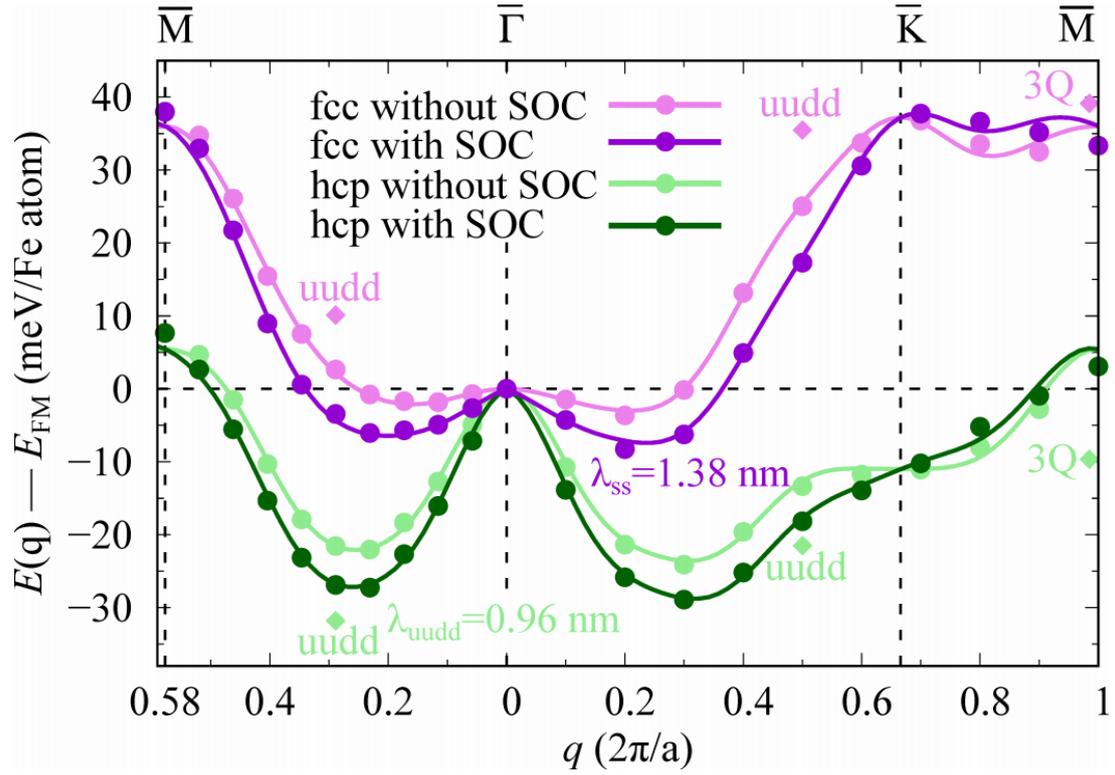

**Figure. 3** Energy dispersion E(**q**) of flat spin spirals along two high-symmetry directions ($\overline{\Gamma KM}$ and $\overline{\Gamma M}$) for fcc-Pd/Fe/Re(0001) (red) and hcp-Pd/Fe/Re(0001) (green). The light (dark) color is used for spin spirals without SOC (with SOC). The filled circles represent DFT data and the solid lines are the fit to the spin model. The filled diamonds represent the 3Q and *uudd* states at the **q** points corresponding to the single-Q states.

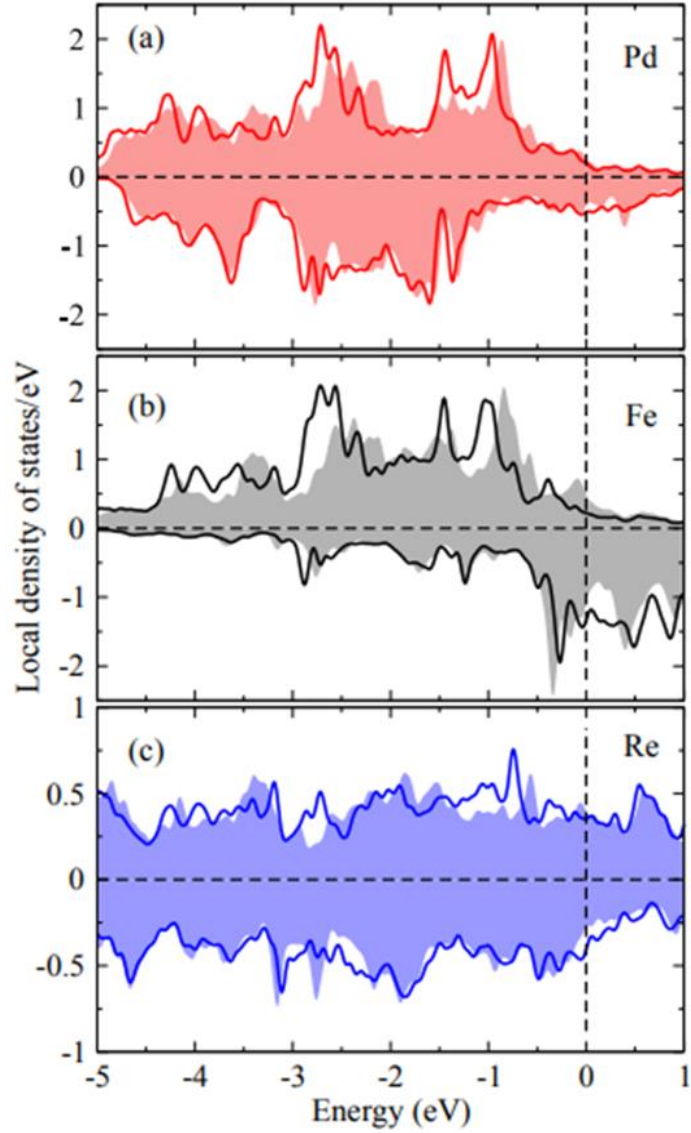

**Figure 4.** Spin-polarized local density of states (LDOS) of (a) 4$d$ Pd, (b) 3$d$ Fe and (c) 5$d$ Re in the FM state. Filled LDOS represent hcp-Pd/hcp-Fe/Re(0001) and solid lines represent fcc-Pd/hcp-Fe/Re(0001).